# Scaling Behaviors of Graphene Nanoribbon FETs: A Three Dimensional Quantum Simulation Study

Yijian Ouyang*, Youngki Yoon*, and Jing Guo†

Department of Electrical and Computer Engineering

University of Florida, Gainesville, FL, 32611

## ABSTRACT

The scaling behaviors of graphene nanoribbon (GNR) Schottky barrier field-effect transistors (SBFETs) are studied by solving the non-equilibrium Green's function (NEGF) transport equation in an atomistic basis set self-consistently with a three-dimensional Poisson equation. The armchair edge GNR channel shares similarities with a zigzag CNT, but it has a different geometry and quantum confinement boundary condition in the transverse direction. The results indicate that the *I-V* characteristics are ambipolar and strongly depend on the GNR width because the bandgap of the GNR is approximately inversely proportional to its width, which agrees with recent experiments. A multiple gate geometry improves immunity to short channel effects, but it offers smaller improvement than it does for Si MOSFETs in terms of the on-current and transconductance. Reducing the oxide thickness is more useful for improving transistor performance than using a high-κ gate insulator. Significant increase of the minimal leakage current is observed when the channel length is scaled below 10 nm because the small effective mass facilitates strong source-drain tunneling. The GNRFET, therefore, does not promise extending the ultimate scaling limit of Si MOSFETs. The intrinsic switching speed of a GNR SBFET, however, is several times faster than that of Si MOSFETs, which could lead to promising high speed electronics applications, where the large leakage of GNR SBFETs is of less concern.

* These authors contributed equally to this work

† e-mail: guoj@ufl.edu

## I. Introduction

The continuous scaling of silicon transistors has been the driving engine for the exponential growth of digital information processing systems over the last decades. The Si transistor in production today is below the 100 nm scale and has entered the nanoelectronics regime. With the scaling limit of Si field-effect transistors (FETs) in sight, a large group of emerging research devices are being extensively studied [1]. Among them, carbon-based nanostructure FETs are the forerunners due to their exceptionally excellent carrier transport properties [2]. Carbon nanotube (CNT) FETs with excellent device performance have been demonstrated [3]. With promising progress on fabricating and patterning a graphene layer, graphene electronics has been a topic of strong research interests [4-5]. A narrow strip of graphene, graphene nanoribbon (GNR), can be either metallic or semiconducting, depending on its structure [6]. An exceptionally high mobility (~10,000 $cm^2/V\text{-}s$) of graphene and GNRs has been demonstrated experimentally [4] and theoretically [7], which leads to the promise of near ballistic transport in a nanoscale GNRFET. The channel geometry of a GNRFET can be defined by lithography, which offers potentially better control over a CNTFET. The concept of all graphene circuits, in which GNRFETs are connected by metallic GNR interconnects, has been proposed [8]. Quite recently great progress has been achieved in fabricating graphene filed-effect devices [9, 10]. A recent theoretical study assessed the performance limits of GNRFETs, but it is based on a semiclassical transport model coupled to a simple treatment of self-consistent electrostatics [11]. Quantum tunneling effects and electrostatic short channel effects were not treated, which makes it difficult to explore scaling behaviors and ultimate scaling limits of GNR SBFETs, where the tunneling effects cannot be ignored.

In this work, a comprehensive study on the scaling behaviors of GNRFETs is performed by solving an atomistic quantum transport equation based on the non-equilibrium Green's function (NEGF) formalism self-consistently with a three-dimensional (3D) Poisson equation. The dependence of the *I-V* characteristics, transconductance, subthreshold swing, drain induced barrier lower (DIBL) on the channel length are studied and compared for the single gate (SG), double gate (DG), and wrapped around gate (WG) geometries. The scaling characteristics of the gate insulator thickness and dielectric constant are explored. The roles of the contact size and Schottky barrier height are examined. The intrinsic delay of the GNRFET is simulated and compared to that projected for the Si FETs at the end of the roadmap.

Transistors with different device structures can operate in different ways. For a conventional MOSFET with heavily doped source and drain extensions, the gate modulates the channel conductance. If the heavily doped semiconductor source and drain are replaced by metal source and drain, Schottky barriers (SBs) form between the contacts and the channel, and a SBFET is obtained. The transistor behavior above the threshold is achieved by modulating the tunneling current through the SBs at the two ends of the channel. In this study, we focus our attention on GNR SBFETs with metal source and drain contacts [9, 10].

## II. Approach

*Device Structure:* We simulated GNRFETs with three different gate geometries at room temperature ($T$=300 K) to explore the effect of gate geometry on the performances of GNRFETs. Fig. 1a shows SG GNRFET, which has the advantage of easy fabrication but is not optimized for good gate control and suppression of short channel effects. Fig. 1b is DG GNRFET, which sandwiches a graphene ribbon between two gates. Fig. 1c shows the cross section of a WG GNRFET (in the plane normal to the channel direction), and the GNR is surrounded by the gate. The WG GNRFET is most challenging for fabrication, but it offers ideal gate control. The nominal device parameters are as follows. The $SiO_2$ gate oxide thickness is $t_{ox}$=2 nm and the relative dielectric constant is $\varepsilon_r$=4. The GNR channel has armchair edges, as shown in Fig. 2. The ribbon index $N$ denotes the number of carbon atom dimmer lines, following the definition in Ref. [6]. The $N$=12 armchair edge GNR channel has a width of ~13.5 Å, which results in a bandgap of $E_g$≈0.83 eV. The channel length is $L_{ch}$=20 nm. The metal source/drain is directly attached to the GNR channel and the Schottky barrier height between the source/drain and the channel is a half of the GNR band gap, $\Phi_{Bn} = \Phi_{Bp} = E_g / 2$. The flat-band voltage is zero. A power supply voltage of $V_{DD}$=0.5 V is used. The nominal device parameters are varied to explore different scaling issues.

*Quantum Transport:* The DC characteristics of GNRFETs are simulated by solving the Schrödinger equation using the non-equilibrium Green's function (NEGF) formalism self-consistently with a 3D Poisson equation [12]. Ballistic transport is assumed [13]. A tight binding Hamiltonian with a $p_z$ orbital basis set is used to describe an atomistic physical observation of the GNR channel. One $p_z$ orbital per atom is enough for the atomistic physical description since s, $p_x$, and $p_y$ are far from the Fermi level and do not play important roles for carrier transport. A $p_z$ orbital coupling parameter of 3 eV is used and only the nearest neighbor coupling is considered. In this study, we use a real space approach rather than a mode space approach. The mode space approach is computationally efficient but it requires uniform potential in transverse direction. To treat possibly varying potential in channel width direction, we use a real space approach. The size of the Hamiltonian is $N \times N$, where $N$ is the number of carbon atoms in the channel. The retarded Green's function of the device is computed as,

$$G(E) = [(E + i0^+)i - H - U - \Sigma_1 - \Sigma_2]^{-1}, \tag{1}$$

where $H$ is the tight binding Hamiltonian matrix of the GNR channel, $U$ is the self-consistent potential matrix determined by the solution of a 3D Poisson equation, and $\Sigma_1$ and $\Sigma_2$ are the self-energies of the metal source and drain contacts, as shown in Fig. 2. The charge density can be computed as,

$$Q_i(x) = (-q)\int_{-\infty}^{+\infty} dE \cdot \mathrm{sgn}[E - E_N(x)]\{D_1(E,x) f(\mathrm{sgn}[E - E_N(x)](E - E_{F1})) + D_2(E,x) f(\mathrm{sgn}[E - E_N(x)](E - E_{F2}))\}, \tag{2}$$

where $q$ is the electron charge magnitude, $\mathrm{sgn}(E)$ is the sign function, $E_{F1,F2}$ is the source (drain) Fermi level, and $D_{1,2}(E,x)$ is the local density of states due to the source (drain) contact, which is computed by the NEGF method. The charge neutrality level [14], $E_N(x)$, is at the middle of band gap because the conduction band and the valence band of the GNR are symmetric.

*Three-Dimensional Electrostatics:* The self-consistent potential is computed from the charge density and the electrode potentials using the Poisson equation,

$$\nabla \cdot [\varepsilon \nabla U(\vec{r})] = qQ(\vec{r}) \tag{3}$$

where $U(\vec{r})$ is the electron potential energy which determines the diagonal entry of the potential energy matrix in Eq. (1), $\varepsilon$ is the dielectric constant, and $Q(\vec{r})$ is the charge density. Because the electric field varies in all dimensions for the simulated device structure, a 3D Poisson equation is numerically solved using the finite element method (FEM). The FEM has the advantage to treat complex device geometries and boundaries between different dielectric materials.

*Device performance metrics:* The source-drain current is computed using the NEGF method once the self-consistency between the NEGF transport equation and the Poisson equation is achieved. The on current, off current and minimal leakage current can be extracted from the simulated *I-V* characteristics.

We compute the intrinsic delay as $\tau = (Q_{on} - Q_{off}) / I_{on}$, where $Q_{on}$ and $Q_{off}$ are the total charge in the channel at on state and off state, respectively, and $I_{on}$ is the on current [15]. The total charge in the channel is calculated by $Q_{ch} = \int_0^{L_{ch}} Q_i(x)dx$, where $Q_i(x)$ is the charge density as a function of channel position. The off state is chosen as the state at which a GNRFET delivers the minimum source-drain current, $V_G=V_D/2$, as discussed later. It can be shifted to $V_G = 0$ if a proper gate work function (and therefore, the threshold voltage) could be achieved. The calculation of the intrinsic delay assumes zero parasitic capacitance. The parasitic capacitance between electrodes impairs the performance of nanoscale transistors in practice. The intrinsic delay, therefore, indicates the upper limit of the switching speed.

## III. Simulation results

We first compare the switching on and switching off characteristics of GNRFETs with different gate geometries. The switching on characteristics are described by simulating the transconductance, and the switching off performance and the immunity of the short channel effects are studied by simulating the subthreshold swing and DIBL. Fig. 3 plots the transconductance, $g_m = \partial I_D / \partial V_G$, which is the ratio of current variation to the gate voltage variation at on state, as a function of the channel length for three types of gate geometries with nominal device parameters. The following observations were made. First, the transconductance remains approximately constant for $L_{ch}$>15 nm. The reason is that the current does not depend on the channel length in the ballistic transport regime, which is different from the diffusive transport regime. Second, as the channel length scales down and approaches to 10 nm, the transconductance decreases because the gate has worse control over the channel due to electrostatic short channel effect, which is conspicuous for SG GNRFET. Third, the transconductances of DG and WG GNRFET are 58 % and 85 % larger than that of SG GNRFET at $L_{ch}$=20 nm. Using a DG structure increases the transconductance but does not double transconductance. In contrast, DG Si

FETs have been actively studied for the promise of the on current advantage over their SG counterpart. The reason is that for a transistor operating at the conventional MOSFET limit (in which the semiconductor capacitance is much larger than the gate insulator capacitance), a double gate structure leads to a two times larger gate insulator capacitance, and therefore, two times larger total gate capacitance and transconductance. For a transistor operating at the quantum capacitance limit (in which the gate insulator capacitance is much larger than the semiconductor capacitance of the channel) [16, 17], the total gate capacitance is limited by the semiconductor capacitance and is independent of the gate insulator capacitance. A double gate geometry does not lead to an improvement of the total gate capacitance and the transconductance. The GNRFET has a one-dimensional channel with a monolayer of carbon atoms. Its small semiconductor capacitance (due to a low density of states) makes it operate closer to the quantum capacitance limit than Si FETs. The advantage of using multiple gate structures, therefore, is smaller.

Fig. 4a shows the variation of subthreshold swing with different channel length. The subthreshold swing, $S$, is calculated by $\Delta V_G / \Delta \log I_D$ at subthreshold region. With 10 nm GNR channel, the gate electrostatic control is not sufficient to keep the subthreshold swing as the longer channel devices. The subthreshold swing, however, remains approximately constant beyond $L_{ch}$=15 nm for all gate geometries. The gate geometry dependence of the subthreshold swing for the nominal device with $L_{ch}$=20 nm is not strong, and the variation of $S$ is less than 6 %. The advantage of multiple gate geometry for the immunity to the short channel effects, however, is more obvious at a shorter channel length. If we choose $S$=100 mV/decade as the criterion (the dotted line in Fig. 4a), the subthreshold swing of SG GNRFET at $L_{ch}$=10 nm does not meet it. The criterion, however, can be met by using a DG geometry (S~96 mV/decade) or a WG geometry (~91 mV/decade).

Fig. 4b shows DIBL vs. channel length. DIBL is a feature of short channel effects and can quantitatively be expressed as $\Delta V_t / \Delta V_D$, where $V_t$ is the threshold voltage. At a channel length of 10 nm, the short channel effect is severe and drain voltage affects the barrier at the beginning of the channel a lot. When channel length increases, DIBL, however, decreases drastically and it remains approximately constant beyond $L_{ch}$=20 nm. For the nominal device with $L_{ch}$=20 nm, DIBL has nearly no difference among all three gate geometries. If we specify 100 mV/V as the criterion (the dotted line in Fig. 4b), at $L_{ch}$=10 nm, DIBL of SG GNRFET is much larger than the criterion, whereas that of DG or WG GNRFET is just above or even less than the criterion, which means GNRFETs with multiple gate geometries can extend the scaling down of channel length more than SG GNRFET. The observation is consistent with that of Fig. 4a.

We next study the effect of power supply voltages. The $I_D$ vs. $V_G$ characteristics for the SG nominal device is plotted in Fig. 5a. As the Schottky barrier height is a half of the bandgap, the minimum currents occur at a gate voltage of $V_{G,min}$=1/2 $V_D$, at which the conduction band bending at the source end of the channel is symmetric to the valence band bending at the drain end of the channel, and the electron current is equal to the hole current. The condition of achieving the minimal leakage current is the same as CNT SBFET with middle gap SBs [18]. Increasing the drain voltage leads to an exponential increase of the minimal leakage current, which indicates the importance of proper designing of the power supply voltage for achieving sufficiently small leakage current.

The minimal leakage bias point can be shifted to $V_G \approx 0$ in order to achieve a small off current (at $V_G=0$) by properly designing the transistor threshold voltage (e.g., by gate work function design). We, therefore, obtain the off state characteristics at $V_{G,min}=V_{DD}/2$ and the on state characteristics at $V_G=V_{G,min}+V_{DD}$ from the simulated *I-V* for GNRFETs with a zero flat band voltage. The output characteristics in Fig.5b show typical linear and saturation regimes. When $V_G$ is increased, saturation current is increased due to a larger voltage drop between the gate and the source contact and a larger energy range for carrier injection from the source contact into the channel.

The effect of channel length scaling on SG GNRFET *I-V* characteristic is explored in Fig. 6. At $V_D$=0.5 V, the off current is increased when channel length decreases to a small value. For a channel length of 5 nm, direct tunneling from the source to drain leads to a large leakage current, and the gate voltage can hardly modulate the current. The transistor is too leaky to have appreciable difference between on and off states. For channel length of 10 nm, the on-off current ratio is greatly improved to near 240. Increasing the channel length to 15 nm, the on-off current ratio is further increased to about 1800 due to decreased direct tunneling current at off state. However, after channel length exceeds 15 nm, the increase in on-off current ratio is not significant. This is because the device is simulated at ballistic limit. (The state-of-the-art short channel CNTFETs already operate close to ballistic limit [3].) Further increasing the channel length hardly changes the on current or off current, nor does the on-off current ratio while for conventional MOSFET, increasing the channel length will cause the channel resistance to increase proportionally.

The $I_D$-$V_D$ characteristics in Fig. 6b confirm the results in Fig. 6a. For 5 nm and 10 nm long channels, direct tunneling from the source to drain and electrostatic short channel effects are severe, so no decent saturation of $I_D$-$V_D$ characteristics is observed. The devices with extremely short channel operate more like a conductor rather than a transistor at $V_G$=0.75 V. For channel length of 15 to 25 nm, typical transistor $I_D$-$V_D$ curves can be obtained and the difference in length results in only a little change in $I_D$-$V_D$ curves. For DG and WG devices, the qualitative results are similar. Compared to single gate device, the channel length can be scaled shorter before the significant loss of on-off current ratio occurs.

Recent experiments studied the dependence of GNR channel resistivity on channel width [9, 10], which shows decreased resistivity with increased width. The scaling of channel width is achieved by changing the index $N$ of the armchair edge GNR channel. Similar to the CNTs, there are semiconducting and metallic GNRs and only semiconducting ribbons are relevant here. In Fig. 7a, the $I_D$-$V_G$ characteristics for different channel widths are plotted. As the channel width is increased, both the off current and on current increase. A GNR with a larger channel width has a smaller bandgap [7], which results in a smaller Schottky barrier height in our simulation and consequently both on and off currents are increased. There is a trade-off between on-off current ratio and on current. For a $N$=9 channel, the on-off current ratio is over 77 thousand while the on current is less than 1 μA. For a $N$=24 armchair edge ribbon channel, although the on current is 9.5 μA, the on-off current ratio is only 8.6, which is close to the recently demonstrated GNR filed-effect device [10]. But unlike the channel length scaling, where extremely short channel device with small on-off current ratio behaves like a conductor, the wide channel device with small on-off current ratio has $I_D$-$V_D$ characteristics with clear linear regime and saturation regime as shown Fig. 7b. Because the channel length here is 20 nm, the gate, therefore, has good electrostatic control over the channel potential profile, and direct tunneling from source to drain is negligible.

The effect of gate oxide dielectric constant is explored. The subthreshold swing and transconductance versus gate oxide dielectric constant are shown in Fig. 8. The increase of gate oxide dielectric constant does not affect much the subthreshold swing and the transconductance. In this study, electrostatic potential in the channel region is carefully calculated by solving the Poisson equation with the Laplace solution combined with particular solution. The Laplace solution for the simulated devices is totally determined by the device geometry and boundary condition, whereas the particular solution is affected by the gate insulator capacitance with its thickness and dielectric constant. The effect of gate dielectric constant on total electrostatic potential is applicable only in the particular solution, but it does not affect much on total electrostatic potential since the magnitude of the particular solution is very small compared to Laplace solution at the quantum capacitance limit [19]. The subthreshold swing and the transconductance, therefore, do not change a lot in the simulated region.

The scaling of the gate oxide thickness is studied next. Fig. 9 shows the subthreshold swing and transconductance versus the gate oxide thickness. The increase of the gate oxide thickness results in worse gate electrostatic control, and hence larger subthreshold swing. A thinner oxide, therefore, is desirable for a larger on-current (due to a larger transconductance) and a larger maximum on-off current ratio (due to a smaller subthreshold swing). As the oxide thickness increases, the role of gate geometry is more important. For example, $g_m$ of the WG GNRFET is about 55 % larger than that of a SG GNRFET for $t_{ox}$=1 nm, but the improvement is 130 % at $t_{ox}$=2.5 nm.

Fig. 10 shows the $I_D$-$V_G$ characteristics of GNR transistor with different Schottky barrier heights. As discussed above, at $V_D$=0.5 V, the minimal leakage current occurs at $V_G$=0.25 V for the Schottky barrier height equal to one half of the bandgap, and the *I-V* characteristic is ambipolar. Because the ambipolar characteristic increases the leakage current and is not preferred for CMOS applications, it is interesting to ask whether the ambipolar characteristic can be suppressed by designing the SB heights. For example, it could be expected that electron transport would be preferred when the SB height for electrons decreases, and the hole transport would be preferred if the SB height for electrons increases. The simulated *I-V* characteristics in Fig. 10, however, indicate that engineering the SB height does not change the qualitative ambipolar feature of the *I-V* characteristics when the gate oxide is thin. The reason is that the gate electrode can screen the field from the source and drain effectively for a thin gate oxide. The Schottky barrier, whose thickness is approximately the same as the gate insulator thickness [20], is nearly transparent. Engineering the SB height, therefore, has a small effect on the qualitative feature of the $I_D$-$V_G$ characteristics. A similar phenomenon has been previously noticed and verified in CNT SBFETs.

The effect of the size of source and drain contacts on transistor *I-V* is studied in Fig. 11. In Fig. 11a, it shows both off current and on current increase as the width of source/drain contact is decreased from 4.8 nm to 1.4 nm while the on-off current ratio keeps approximately constant. To understand the $I_D$-$V_G$ curves, the conduction band profiles at on state ($V_D$=0.5 V and $V_G$=0.75 V) in transport direction are plotted in Fig. 11b. Because the simulation shows little potential variation in the transverse direction of the channel, the potential is obtained by averaging the potential of each atomistic site in the channel width direction. It clearly shows the thickness of Schottky barrier at source end is decreased when the width of contacts is decreased. This is due to the fact that scaling down the width of the contacts can decrease the fringe field from gate to source contact and the penetration length of source electrical field is decreased. The Fig. 11c and 11d show how the height of contacts affects the transistor's $I_D$-$V_G$ characteristics. It shows the same trend when the contact height is changed. The DG and WG transistors basically follow the same qualitative conclusions. Compared to the SG FET, the contact size has smaller effect on the DG and WG GNRFETs. Because the GNR channel is sandwiched or wrapped around by the gate, the field lines from the source and drain contacts are better screened by the gate electrodes, and hence the source and drain contact geometry has less impact. Generally, small size and low dimensional contacts not only improve the DC performance by increasing the on current, but also improve the AC performance by decreasing the parasitic capacitance between the electrodes, which leads to a higher operation speed.

The intrinsic delay indicates how fast a transistor can switch. In Fig. 12, the intrinsic delays of a GNRFET and a Si MOSFET are compared. The intrinsic delay of the Si MOSFET is projected by ITRS to year 2015 and the gate length $L_g$ will be shrunk down to 10 nm [1]. The delay data ($C_g V / I$) in the ITRS roadmap are computed using the total gate capacitance $C_g$ that includes the parasitic capacitance, as shown by the dashed line with circles in Fig. 12. In order to perform a fair comparison to the intrinsic delay of the nominal SG GNRFETs (which does not include the parasitic capacitance), we also compute the intrinsic delay of Si MOSFETs by using the ideal gate capacitance values provided by the ITRS roadmap instead of the total gate capacitance values, as shown by the dashed line. The results indicate that the intrinsic delay of GNRFET is approximately a factor of 3.5 smaller than that of Si MOSFETs at $L_g$=10 nm. The major reason is that the GNR has a larger band-structure limited velocity for the first subband, which dominantly contributes to the conduction in low voltage operation.

## IV. Conclusions and Discussions

In this work, a comprehensive study on the scaling characteristics of GNR SBFET is performed by solving quantum transport equation with self-consistent electrostatics. Due to the geometry of GNR, we solved Poisson equation for a 3D electrostatics. With possibly varying potential in the width direction of GNR channel, an atomistic tight binding Hamiltonian in real space representation is used in transport equation. Such approach can be readily extended to treat GNR channel with edge irregularity and defects where a mode space approach is not applicable.

Because both GNR and CNT are one-dimensional nanostructures derived from graphene (with a bandgap created by quantum confinement in the width or circumferential direction), the scaling characteristics of GNR SBFETs show some similarities with CNT SBFETs. But the two types of transistors have some important difference. First, GNRFETs have different channel geometry leading to a different gate electrostatics. A precise patterning technique could potentially lead to better control of defining channel material than CNTFETs. Second, due to the different quantum confinement in the transverse direction, an armchair edge GNR channel does not have valley degeneracy [11], which results in a smaller quantum capacitance. Compared to a CNT channel [21], it benefits even less from multiple gate structure. In addition, the edges of the GNR channel could potentially cause large potential variation in the transverse direction.

The band gap of the GNR channel strongly depends on its width, which significantly affects the on current and off current. The ambipolar *I-V* characteristics can not be suppressed by engineering the Schottky barrier height when the gate insulator is thin (below about 10 nm). Reducing the gate insulator thickness and the contact size results in thinner Schottky barriers, and therefore, larger on current. The intrinsic speed of the GNRFETs is several times faster than Si FETs due to its large carrier velocity and near ballistic transport, and reducing parasitics is essential for benefiting from the fast intrinsic speed. As a new type of transistor, GNRFET, however, is still governed by transistor electrostatics and quantum effects, which imposes a similar ultimate scaling limit as for Si MOSFETs [22].

**Acknowledgement**

The computational facility was made possible from the NSF Grant No. EIA-0224442, IBM SUR grants and gifts, and a DURIP Grant from Army Research Office.

## FIGURES

Fig. 1 Simulated device structures with different gate geometries: (a) single gate (SG), (b) double gate (DG), and (c) cross section of wrapped around gate (WG) graphene nanoribbon (GNR) FET. The channel materials are GNRs. The source and drain contacts are metals, and the Schottky barrier height is a half band gap of the channel material.

Fig. 2 The schematic sketch of an armchair edge $N$=12 GNR channel with the source and drain contacts. The quantities used in the NEGF formalism are also shown.

Fig. 3 The gate transconductance vs. channel length. The solid line is for SG, the dashed line with crosses is for DG, and the dash-dot line with squares is for WG GNRFET. A better gate control results in larger transconductance.

Fig. 4 Channel length dependence of (a) $S$ and (b) DIBL. The solid line is for SG, the dashed line with crosses is for DG, and the dash-dot line with squares is for WG GNRFET. The multiple gate geometries can extend the scaling down of channel length below 10 nm.

Fig. 5 (a) The $I_D$ vs. $V_G$ characteristics for the nominal single gate device at different source-drain voltages. (b) The $I_D$ vs. $V_D$ characteristics for the nominal single gate device at different gate voltages.

Fig. 6 (a) The $I_D$ vs. $V_G$ characteristics for single gate device with different channel lengths at $V_D$=0.5 V. (b) The $I_D$ vs. $V_D$ characteristics for single gate device with different channel lengths at $V_G$=0.75 V (the curve with crosses is for $L_{ch}$ =5 nm; the curve with circles is for $L_{ch}$ =10 nm; the curve with stars is for $L_{ch}$ =15 nm; the solid line without mark is for $L_{ch}$ =20 nm; the dashed line is for $L_{ch}$ =25 nm).

Fig. 7 (a) The $I_D$ vs. $V_G$ characteristics for single gate device with different channel widths at $V_D$=0.5 V. (b) The $I_D$ vs. $V_D$ characteristics for single gate device with different channel widths at $V_G$=0.75 V. The channel length is 20 nm.

Fig. 8 Gate oxide dielectric constant dependence of (a) $S$ and (b) $g_m$. Solid line is for SG, dashed line with crosses is for DG, and dash-dot line with squares is for WG GNRFET.

Fig. 9 Gate oxide thickness dependence of (a) $S$ and (b) $g_m$. A better gate control results in smaller subthreshold swing and larger transconductance.

Fig. 10 The $I_D$ vs. $V_G$ characteristics for single gate device with different Schottky barrier heights (The curves with triangles, stars, circles, crosses and without marks are for the electron Schottky barrier height of 0, 1/8 $E_g$, 1/4 $E_g$, 3/8 $E_g$ and 1/2 $E_g$, respectively).

Fig. 11 (a) The $I_D$ vs. $V_G$ characteristics for single gate device with contact width equal to 1.4 nm, 3 nm, and 4.8 nm, respectively. (b) The conduction band profiles at on state ($V_D$=0.5V and $V_G$=0.75V) in transport direction for source/drain contact width of 1.4 nm, 3 nm and 4.8 nm. $E_{FS}$ and $E_{FD}$ indicate source and drain Fermi levels, respectively. (c) The $I_D$ vs. $V_G$ characteristics for single gate device with contact height equal to 1 nm, 2 nm and 8 nm, respectively. (d) The conduction band profiles in transport direction for source/drain contact height of 1 nm, 2nm and 8 nm.

Fig. 12 The intrinsic delay vs. gate length for a SG GNRFET (the solid line) and a Si MOSFET (the dashed line). The dashed line with circles is the ITRS' projection on transistor delay with parasitic capacitance, and the dashed line is obtained by using ideal gate capacitance without parasitic capacitance for a fair comparison of the intrinsic delay. For GNRFETs, the gate length $L_g$ is equivalent to the channel length $L_{ch}$.

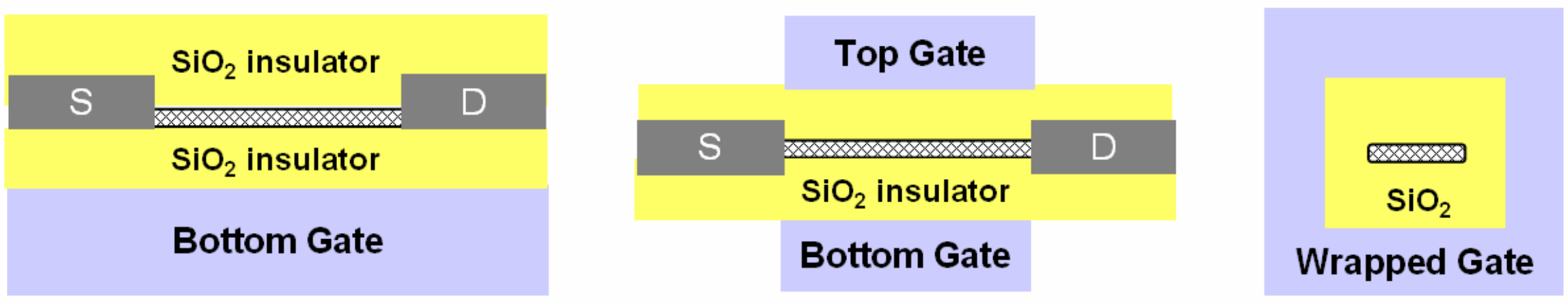

**(a)** **(b)** **(c)**

Fig. 1

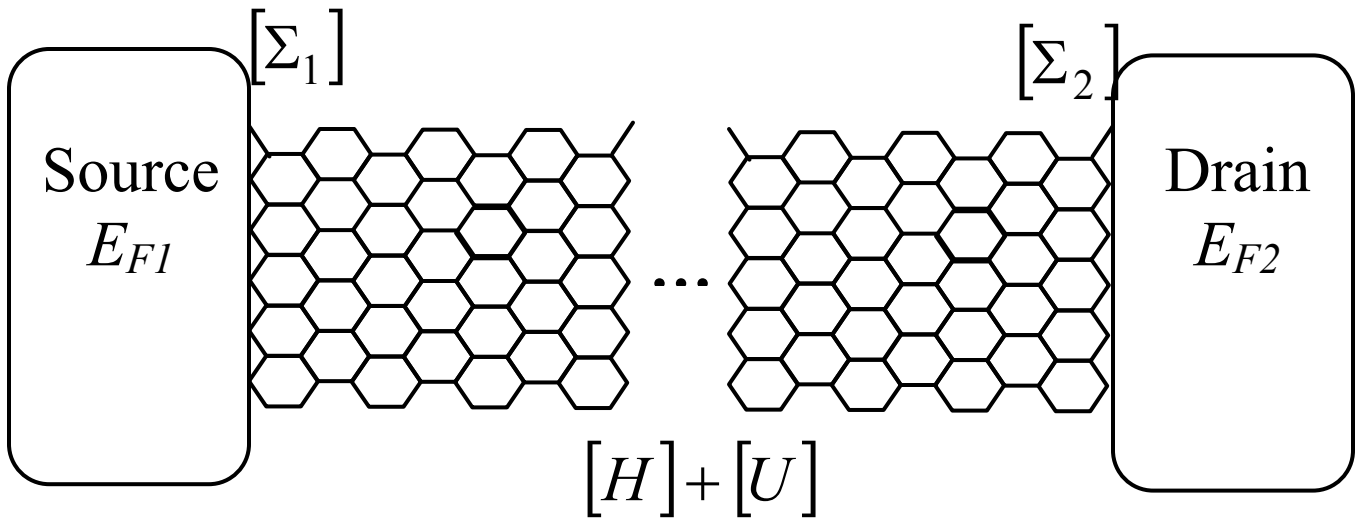

Fig. 2

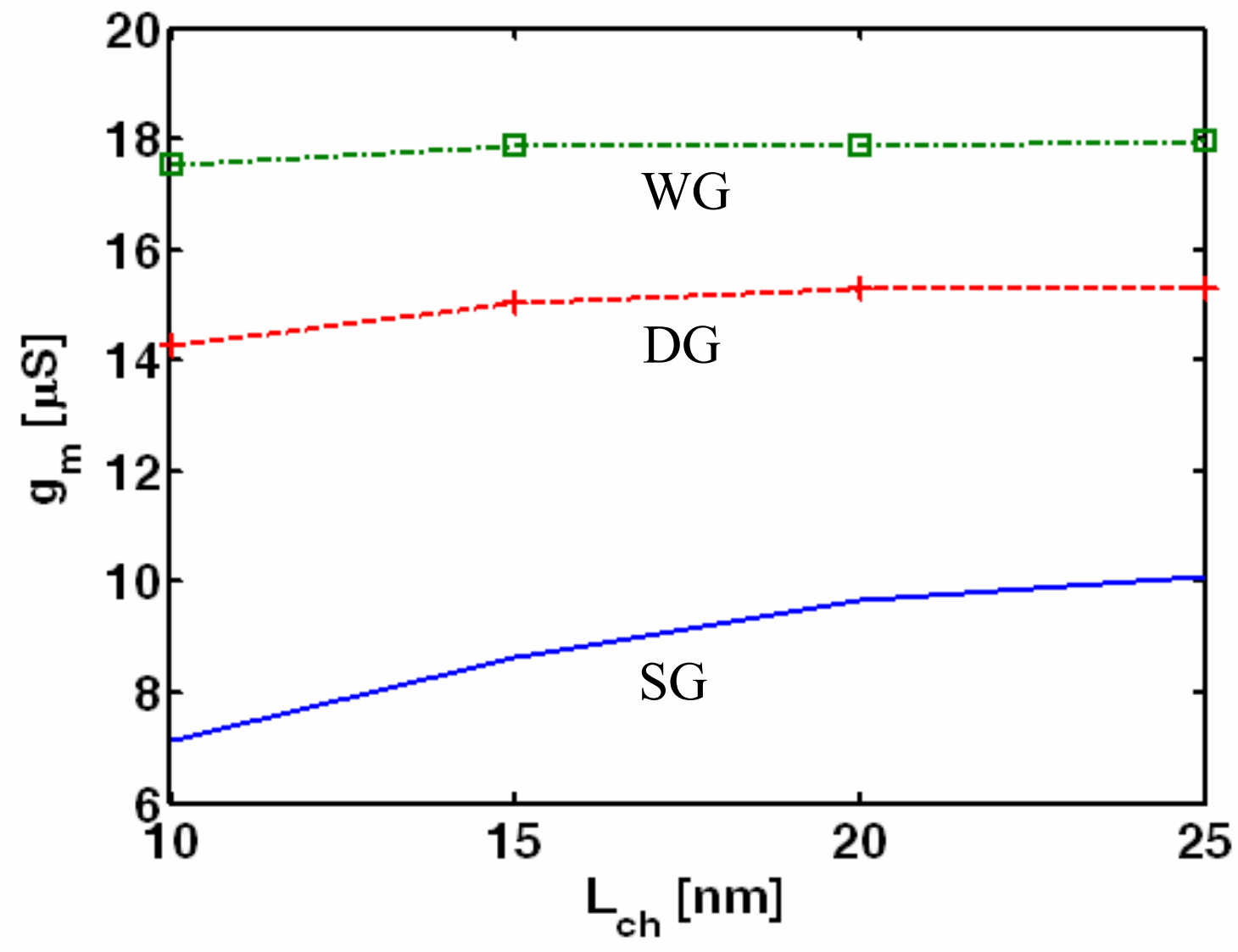

Fig. 3

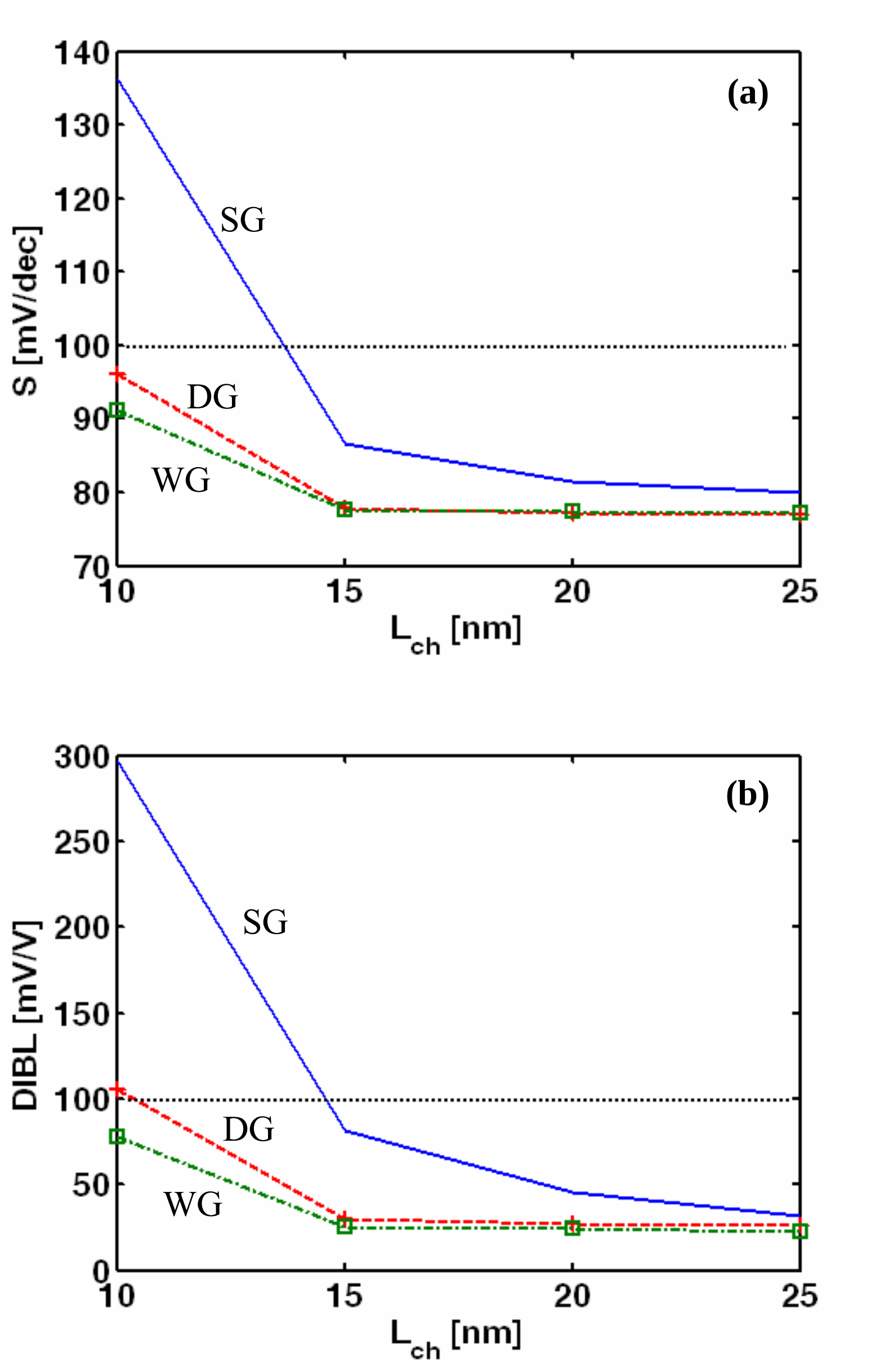

Fig. 4

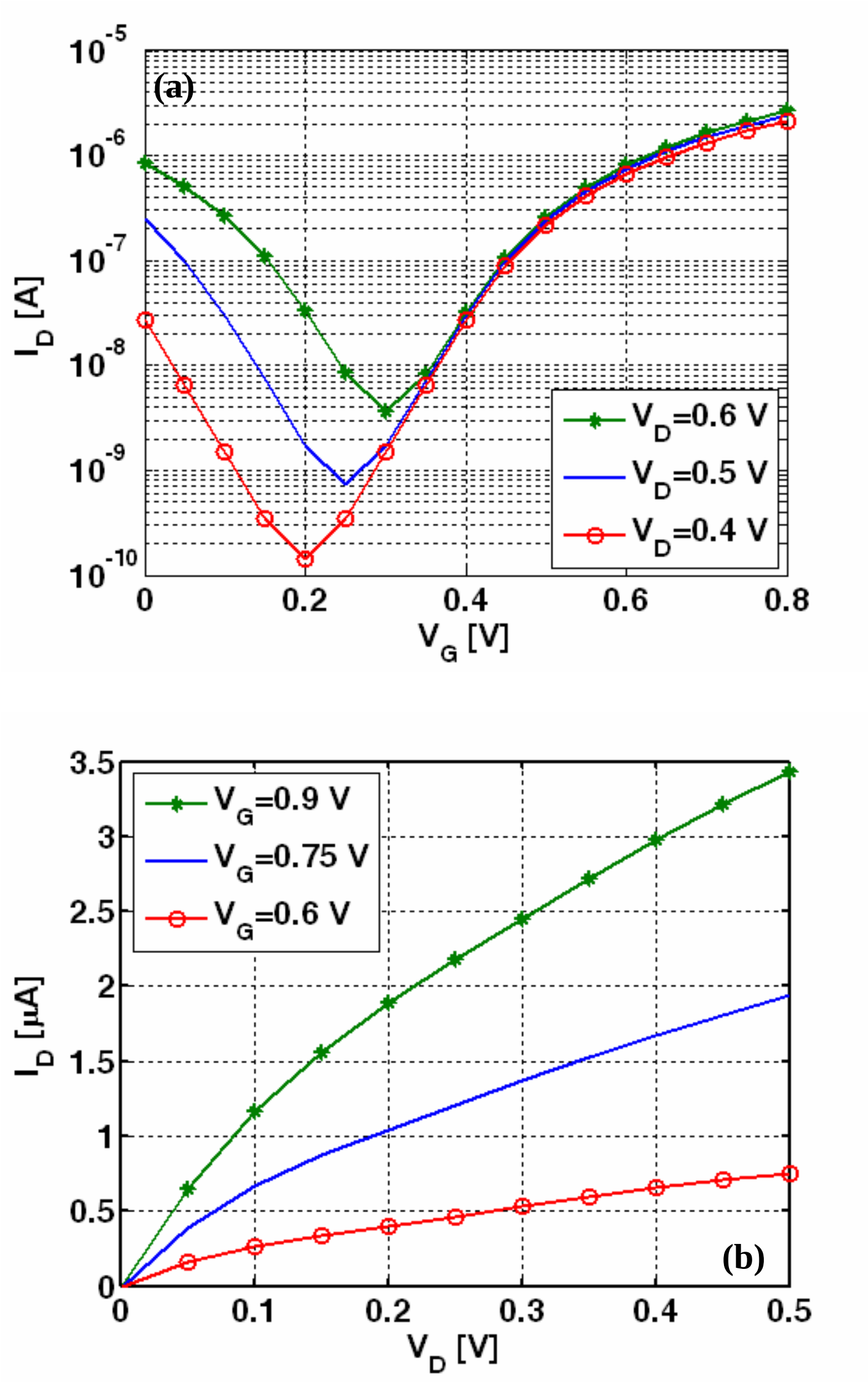

Fig. 5

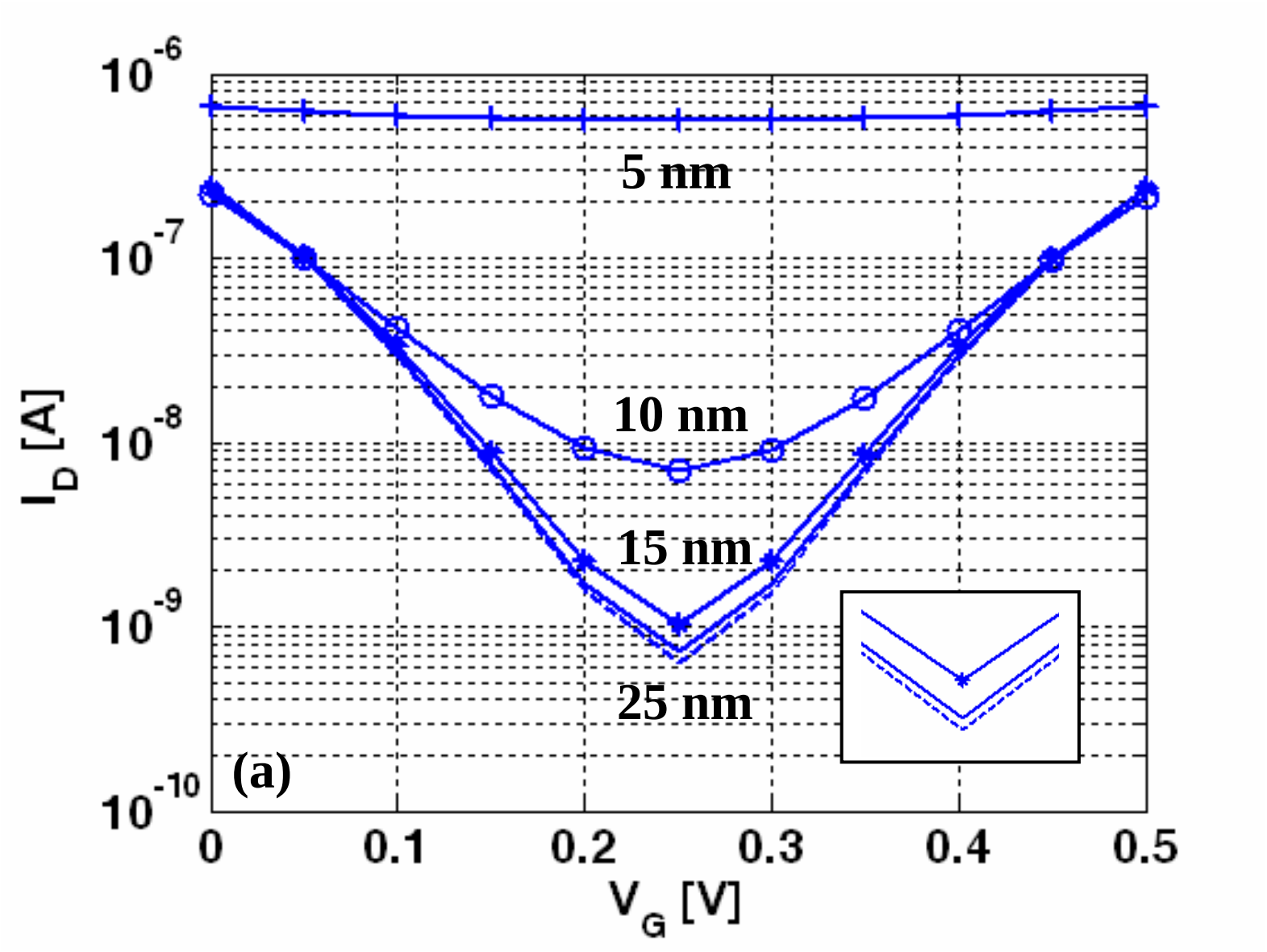

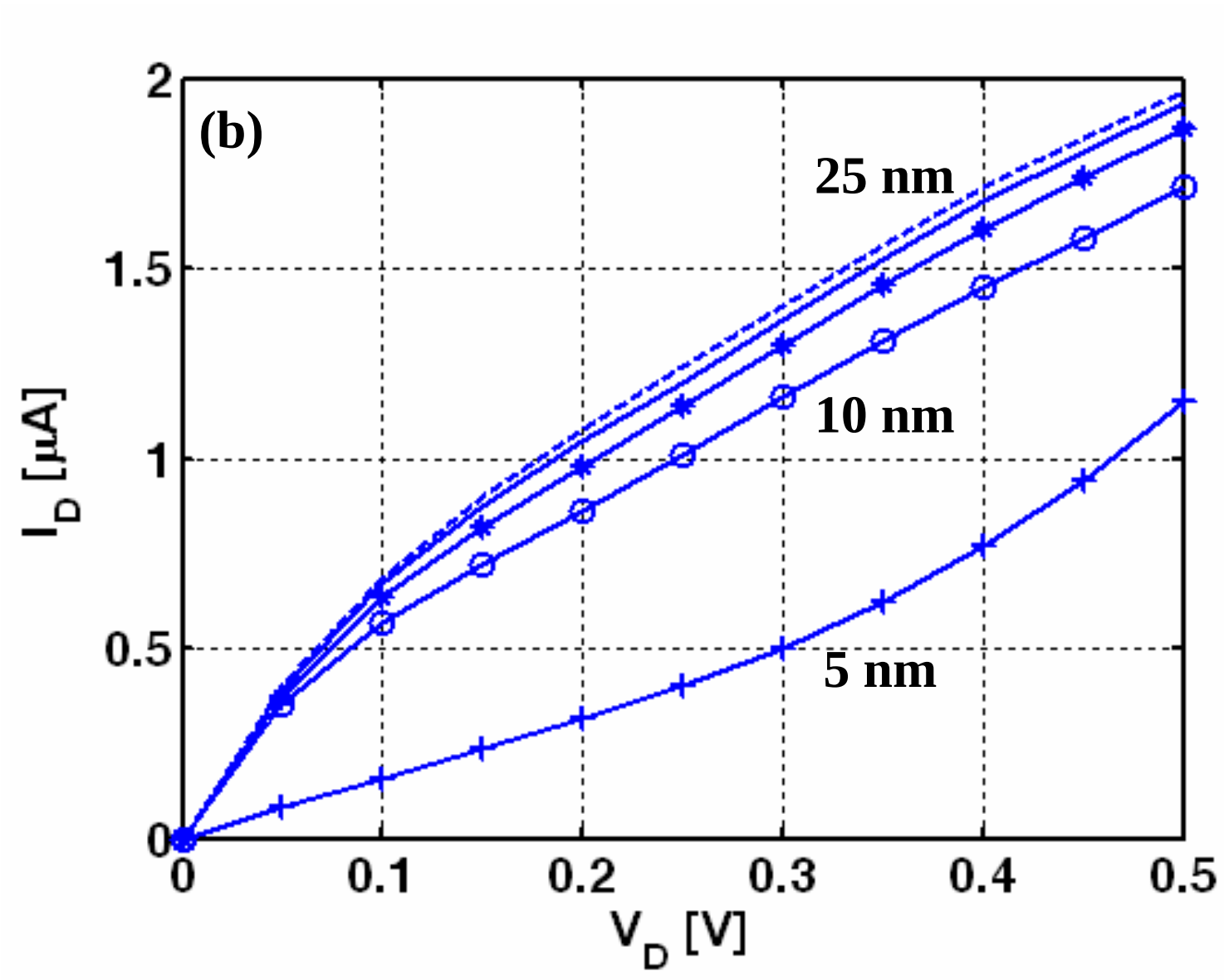

Fig. 6

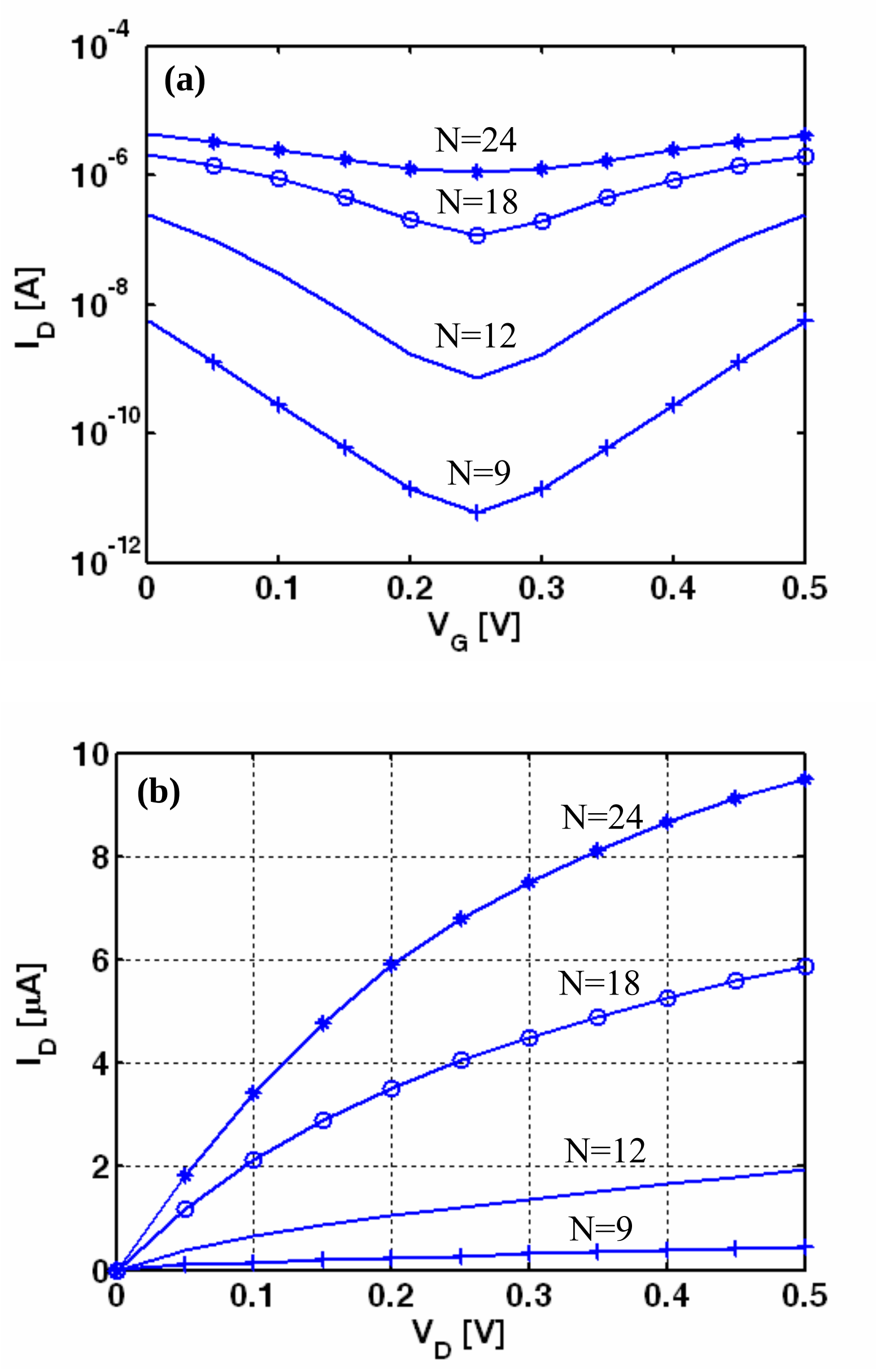

Fig. 7

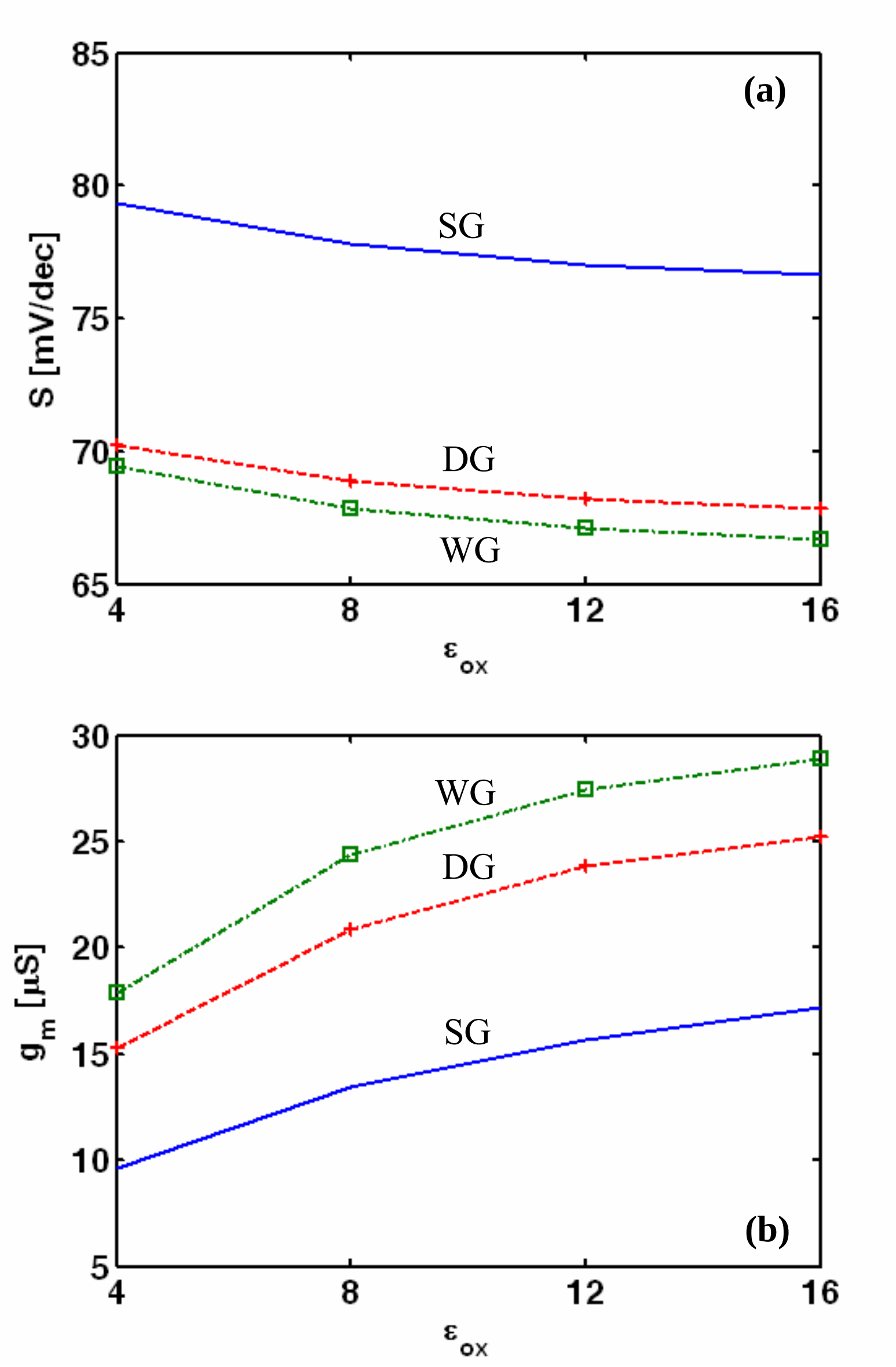

Fig. 8

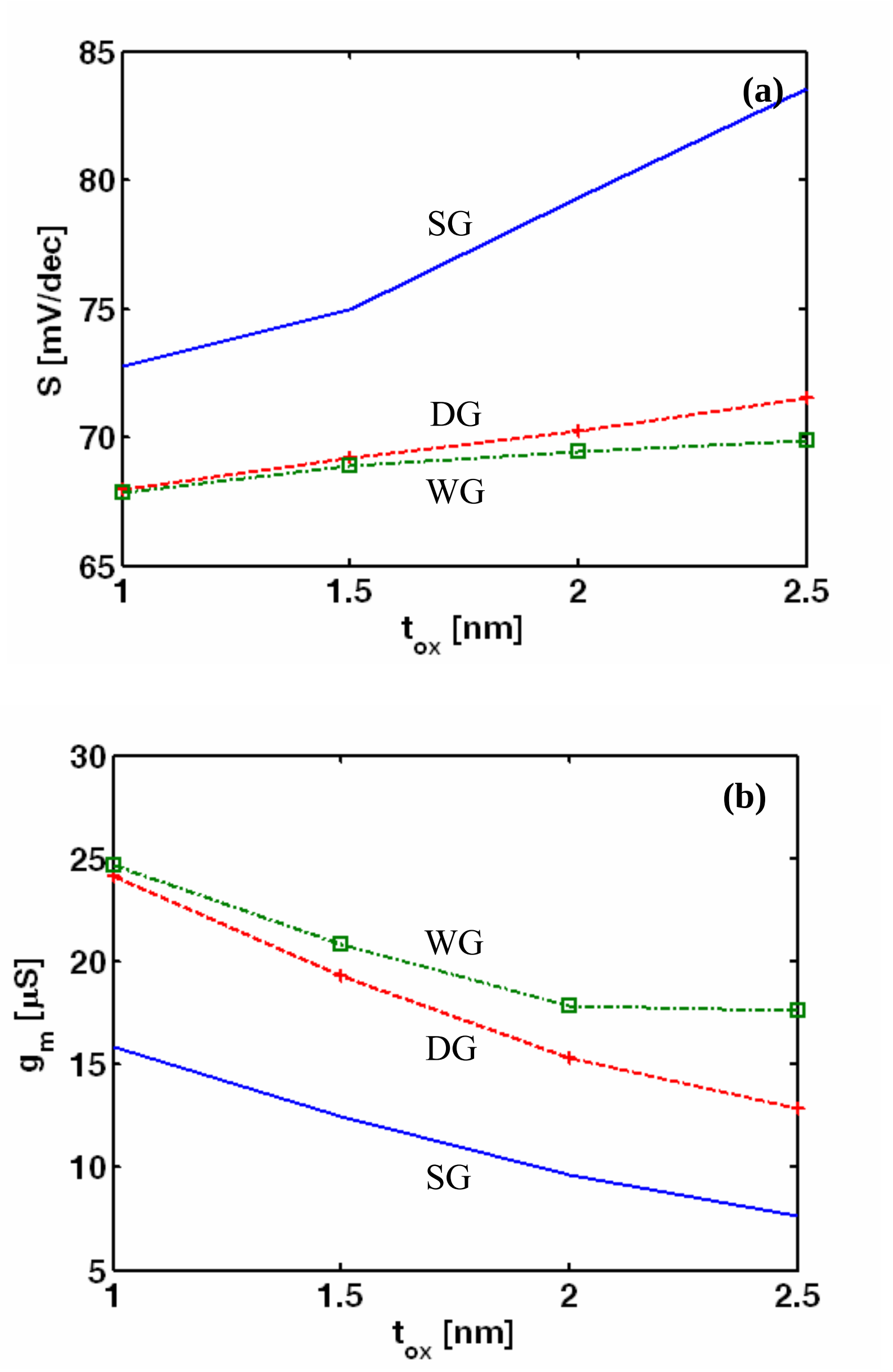

Fig. 9

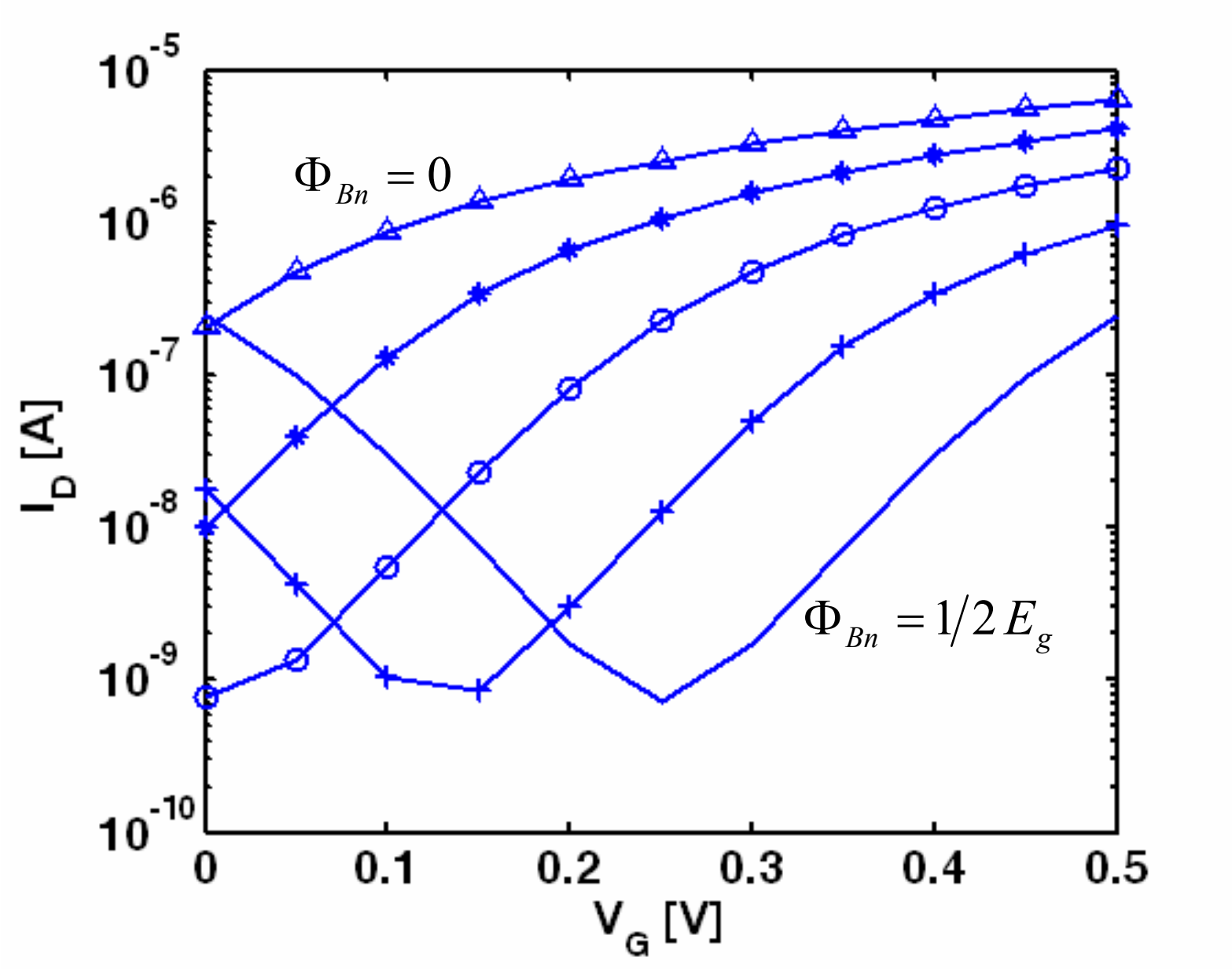

Fig. 10

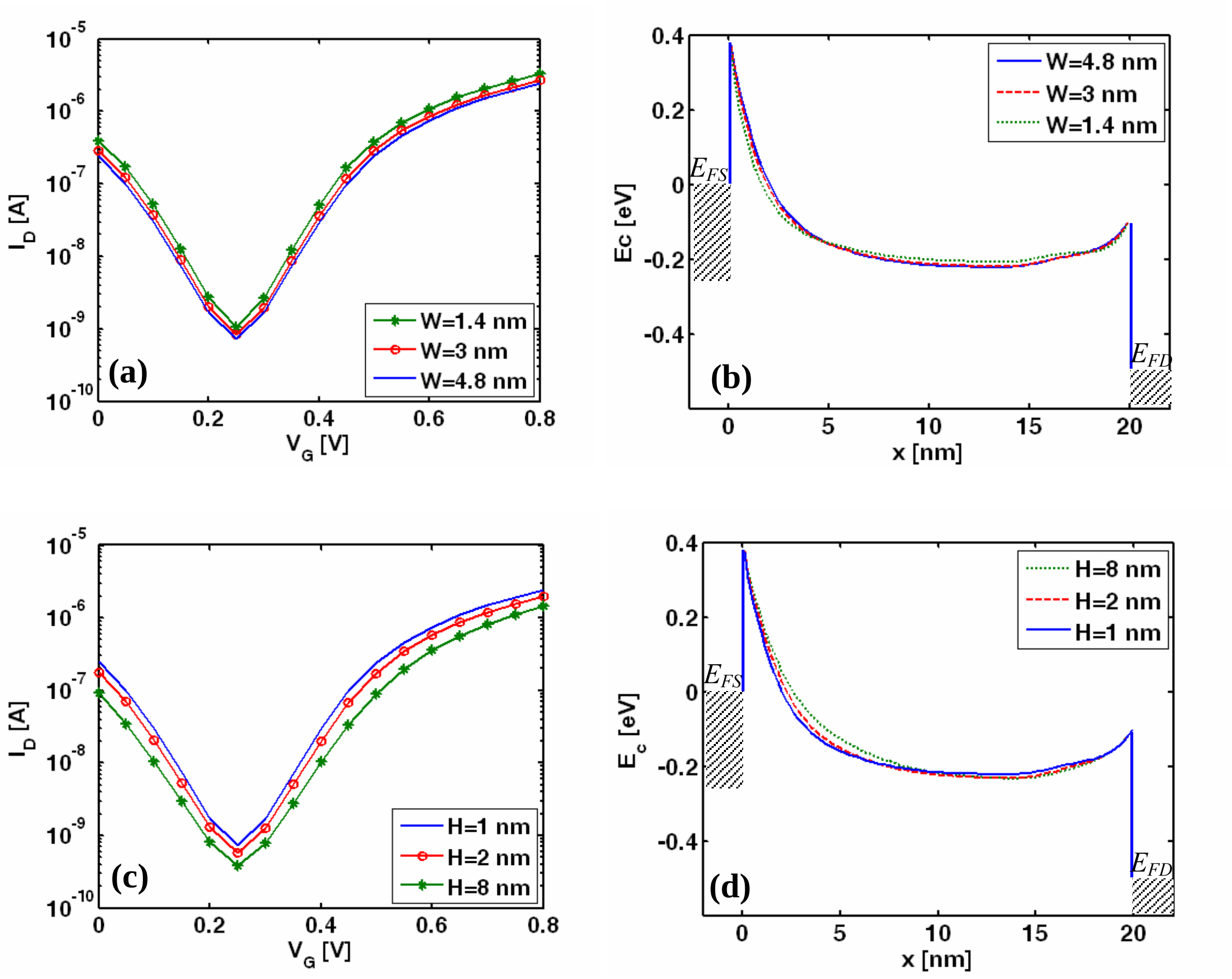

Fig. 11

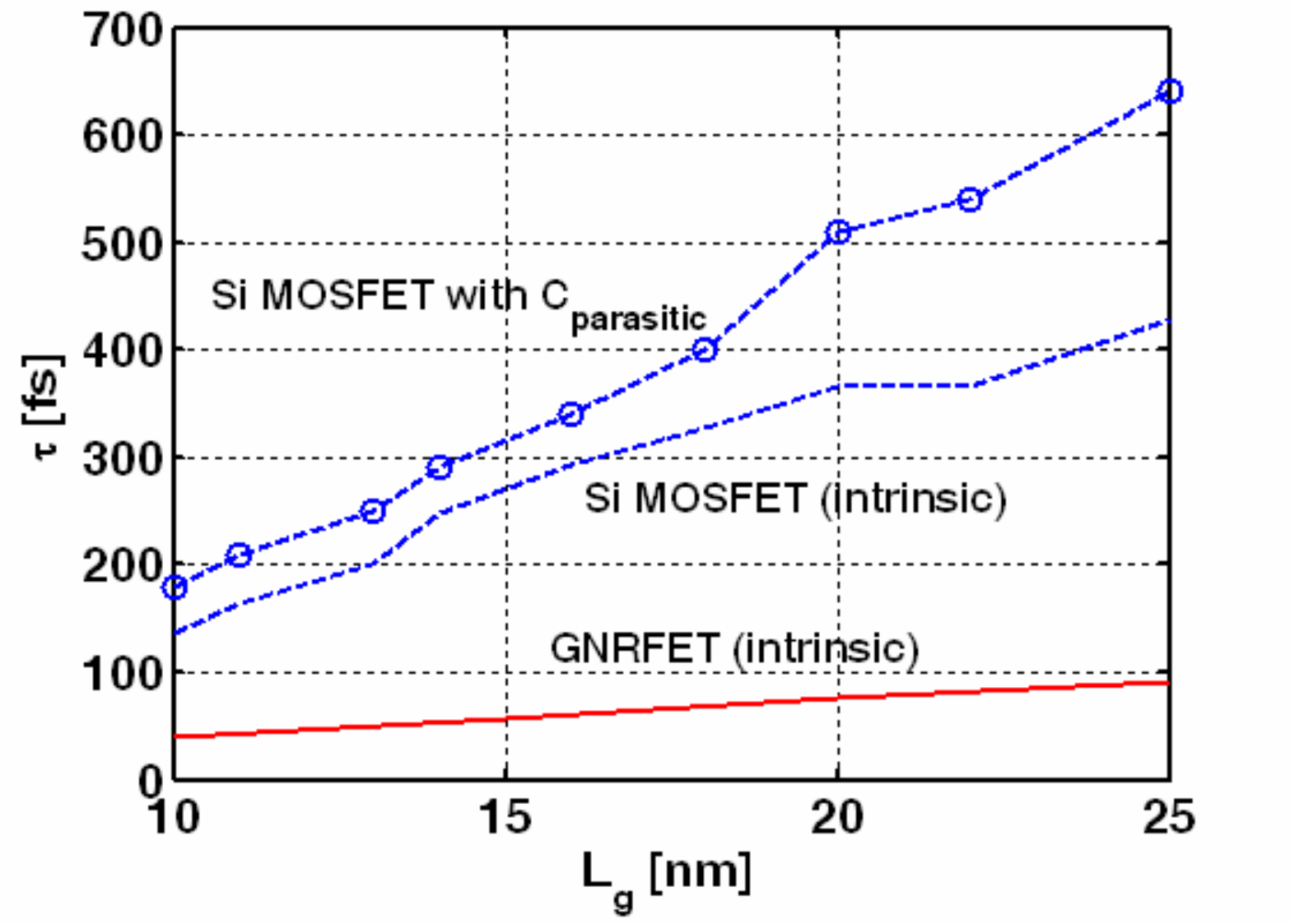

Fig. 12